\documentclass[aps,amsmath,amssymb,showpacs,floatfix,twocolumn,prb,superscriptaddress]{revtex4-1}
\usepackage{graphicx}
\usepackage{dcolumn}
\usepackage{bm}
\usepackage[dvips]{color}

\def\bk{\mathbf{k}}

\begin{document}

\title{Interplane charge dynamics in a valence-bond dynamical mean-field theory\\
       of cuprate superconductors}

\date{\today}

\author{M. Ferrero}
\affiliation{Centre de Physique Th{\'e}orique, CNRS, Ecole
Polytechnique, 91128 Palaiseau Cedex, France}
\author{O. Parcollet}
\affiliation{Institut de Physique Th{\'e}orique, CEA, IPhT, CNRS, URA 2306,
91191 Gif-sur-Yvette, France}
\author{A. Georges}
\affiliation{Centre de Physique Th{\'e}orique, CNRS, Ecole
Polytechnique, 91128 Palaiseau Cedex, France}
\affiliation{Coll{\`e}ge de France, 11 place Marcelin Berthelot, 75005 Paris, France}
\author{G. Kotliar}
\affiliation{Physics Department and Center for Materials Theory, Rutgers University,
Piscataway NJ 08854, USA}
\author{D. N. Basov}
\affiliation{Department of Physics, University of California-San Diego, La Jolla, California 92093, USA}

\begin{abstract}
We present calculations of the interplane charge dynamics in the normal state
of cuprate superconductors within the valence-bond dynamical mean-field theory.
We show that by varying the hole doping, the $c$-axis optical conductivity and
resistivity dramatically change character, going from metallic-like at
large doping to insulating-like at low-doping. We establish a clear connection
between the behavior of the $c$-axis optical and transport properties and the
destruction of coherent quasiparticles as the pseudogap opens in the antinodal
region of the Brillouin zone at low doping. We show that our results are in good agreement
with spectroscopic and optical experiments.
\end{abstract}

\maketitle


\section{Introduction}

Even after many years of investigation, cuprate superconductors continue to be
at the center of intense experimental and theoretical interest.  A key
phenomenon which needs to be understood to reveal the nature of
superconductivity in the cuprates is the onset of strong momentum-space
differentiation in the (hole-) underdoped normal state.  This region of the
phase diagram is characterized by the suppression of quasiparticle excitations
in the antinodal region of the Brillouin zone and the opening of a pseudogap
yielding Fermi arcs, as observed, e.g. in angle-resolved photoemission (ARPES)
experiments.~\cite{damascelli:rmp:2003:1} Other distinctive properties of the
cuprates are seen in the extreme anisotropy of the charge
dynamics.~\cite{ito:nature:1991:1, nakamura:prb:1993:1,
takenaka:prb:1994:1,basov:rmp:2005:1} The in-plane ($\rho_{ab}$) and interplane
resistivity ($\rho_c$) have different temperature dependence. While $\rho_{ab}$
decreases with decreasing temperatures at all doping levels, $\rho_c$ is
sensitive to the doping. Its behavior is essentially insulating at low doping,
semi-metallic at intermediate doping and metallic in the overdoped regime. The
peculiar properties of the interplane charge dynamics have also been observed
in optical studies.~\cite{basov:rmp:2005:1, tamasaku:prl:1992:1,
cooper:prb:1993:1, homes:prl:1993:1, schutzmann:prl:1994:1,
tamasaku:prl:1994:1, puchkov:jpcm:1996:1, basov:prl:1996:1, puchkov:prl:1996:1}
They reveal that the $c$-axis optical conductivity $\sigma_c(\Omega)$ shows no
sharp Drude peak at low frequencies $\Omega$ in the underdoped normal state.
Instead, $\sigma_c(\Omega)$ has a rather flat frequency-dependence at
temperatures above the pseudogap temperature $T^*$. As the temperature is
reduced below $T^*$, low-energy spectral weight is transferred to higher
energies and $\sigma_c(\Omega)$ displays a gaplike depression as $\Omega
\rightarrow 0$.  Only in the highly overdoped region does $\sigma_c(\Omega)$
show evidence of emerging coherence.~\cite{schutzmann:prl:1994:1}

In this article, we address the interplane charge dynamics of cuprate
superconductors within the valence-bond dynamical mean-field theory (VB-DMFT)
introduced in Refs.~\onlinecite{ferrero:epl:2009:1,ferrero:prb:2009:1}.  Our
results capture many salient features found in experiments for the c-axis
resistivity and optical conductivity. Specifically, we show that the opening of a
pseudogap in the antinodal region explains the incoherent behavior of the
interplane transport.

\begin{figure}[ht!]
\begin{center}
  \includegraphics[width=4.0cm,clip=true]{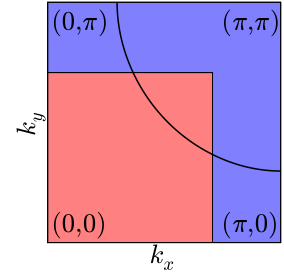}
  \caption{(Color online) The two patches dividing the Brillouin zone. The line
   shows a non-interacting Fermi surface for the dispersion $\epsilon_\bk$ of
   Eq.~\ref{eq:hamilt}. The central (red) patch covers the nodal
   region of the Fermi surface, while the border (blue) patch covers the antinodal region.}
  \label{fg:patches}
\end{center}
\end{figure}

\begin{figure*}[Ht!]
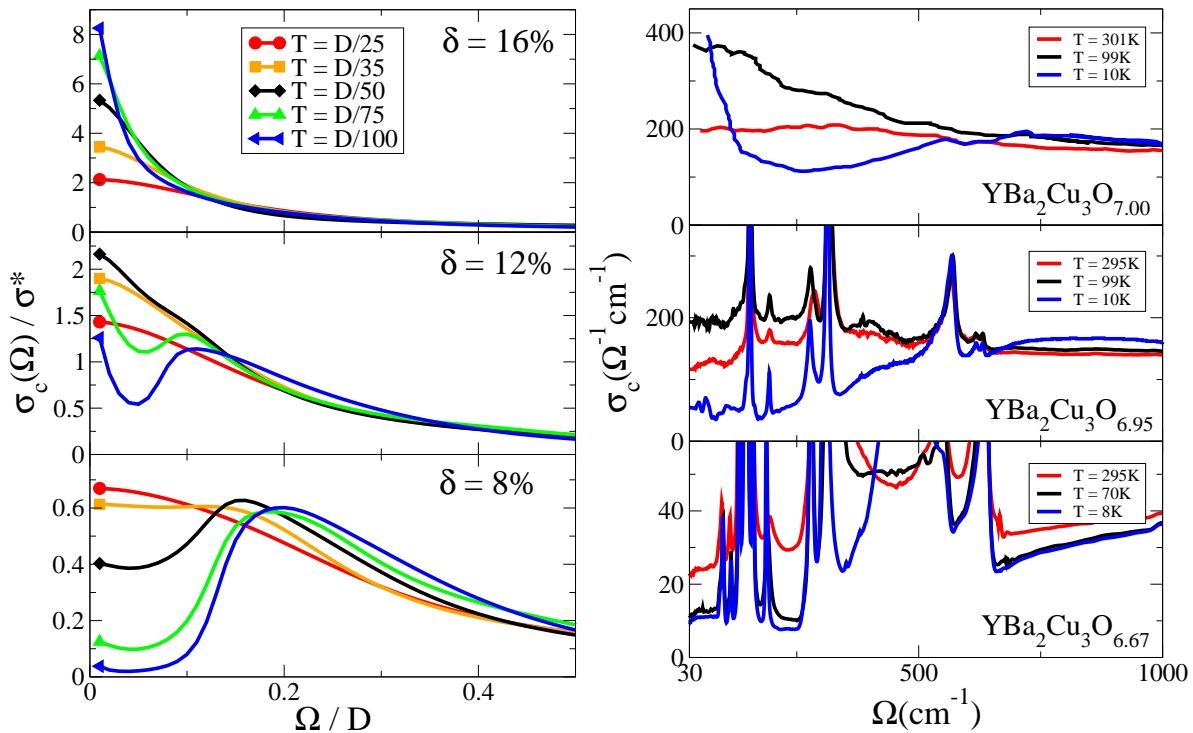

\begin{center}
  \includegraphics[width=7.6cm,clip=true]{fig2a}
  \includegraphics[width=8cm,clip=true]{fig2b}
  \caption{(Color online) Left panel: The $c$-axis optical conductivity $\sigma_c(\Omega)$
  calculated within VB-DMFT for three doping levels. $\sigma_c$ is displayed in
  units of $\sigma^*$ as defined in the text ($\sigma^*$ is of order 50 $\Omega^{-1} \mathrm{cm}^{-1}$
  for YBa$_2$Cu$_3$O$_y$).
  Frequency is normalized to the half-bandwidth $D\sim 1\rm{eV}\,= 8000\rm{cm}^{-1}$.
  Right panel: Experimental data for the $c$-axis optical conductivity of
  YBa$_2$Cu$_3$O$_y$. The data for YBa$_2$Cu$_3$O$_{7.00}$ is taken from
  Ref.~\onlinecite{schutzmann:prl:1994:1} where the phonon contribution was
  subtracted by fitting to 5 Lorentzian oscillators. The data for
  YBa$_2$Cu$_3$O$_{6.95}$ and YBa$_2$Cu$_3$O$_{6.67}$ are taken from
  Refs.~\onlinecite{laforge:prl:2008:1,laforge:prb:2009:1}.
  }
  \label{fg:sigmac}
\end{center}
\end{figure*}

The VB-DMFT is a minimal cluster extension of the dynamical mean-field
theory~\cite{georges:rmp:1996:1,maier:rmp:2005:1} that aims at describing
momentum-space differentiation together with Mott physics. We apply
it to the Hubbard model on a square lattice defined by the Hamiltonian
\begin{eqnarray}
  \mathcal{H} &=& \sum_{\bk,\sigma} \epsilon_\bk
    c^\dagger_{\sigma,\bk} c^{\phantom{\dagger}}_{\sigma,\bk}
  + U \sum_i n_{i \uparrow} n_{i \downarrow}, \label{eq:hamilt} \\
  \epsilon_\bk &=& -2t [\cos(\bk_x) + \cos(\bk_y)] - 4 t^\prime \cos(\bk_x) \cos(\bk_y),\nonumber
\end{eqnarray}
where $t$ and $t^\prime$ are the nearest and next-nearest neighbor hopping and
$U$ is the onsite Coulomb repulsion.  In the following, we use $U/t=10$ and
$t^\prime/t=-0.3$, which are values commonly used for modeling hole-doped
cuprates in a single-band framework.
This value of $U$ is larger than the critical value for the Mott transition in
the undoped model, so that we deal with a doped Mott insulator.
VB-DMFT is based on an effective two-impurity model embedded
in a self-consistent bath. Within this simple approach
involving only two degrees of freedom, the lattice self-energy
is approximated to be piecewise constant over two patches of the Brillouin zone.
These two patches are shown in Fig.~\ref{fg:patches}. Inspired by the
phenomenology of cuprate superconductors, their shape is chosen
in such a way that the central (red) patch covers the nodal region of the Fermi surface,
while the border (blue) patch covers the antinodal region. With this
prescription, one degree of freedom of the underlying two-impurity model
is associated with the physics of nodal quasiparticles and the other
degree of freedom with antinodal quasiparticles. Thank to the
relative simplicity of the VB-DMFT approach, it is possible to
make very efficient calculations using continuous-time quantum Monte Carlo
and obtain accurate spectral functions $A(\bk,\omega)$.
More details concerning this procedure are given in Appendix~\ref{app:vbdmft}
and in Refs.~\onlinecite{ferrero:epl:2009:1,ferrero:prb:2009:1}.

In the VB-DMFT, the formation of Fermi arcs is described by a selective
metal-insulator transition in momentum
space.~\cite{ferrero:epl:2009:1,ferrero:prb:2009:1} Below a doping $\sim 16\%$,
the degree of freedom describing the antinodal regions becomes insulating,
while that associated to the nodal quasiparticles remains metallic. The
orbital-selective mechanism responsible for the pseudogap has also been
confirmed in studies involving larger
clusters.~\cite{gull:prb:2009:1,werner:prb:2009:1,gull:condmat:2010:1} In
Refs.~\onlinecite{ferrero:epl:2009:1,ferrero:prb:2009:1}, we used the VB-DMFT
to compute tunneling and ARPES spectra in good agreement with experiments.


\section{Interplane optical conductivity}

We first compute the frequency-dependent $c$-axis optical conductivity
$\sigma_c(\Omega)$, given by
\begin{equation}
\begin{split}
  \sigma_c(\Omega)
    & = \frac{2 e^2 c}{\hbar a b} \int \!\! d\omega \, \frac{f(\omega) - f(\omega+\Omega)}{\Omega}  \\
    & \quad \frac{1}{N} \sum_\bk t^2_\perp (\bk) A(\bk,\omega) A(\bk,\Omega+\omega),
\end{split}
\label{eq:sigmac}
\end{equation}
where $f$ is the Fermi function, $A(\bk,\omega)$ the in-plane spectral function,
$N$ the number of lattice sites, $e$ the electronic charge, $a,b$
the in-plane lattice constants, $c$ the interplane distance and
$t_\perp(\bk) = t_0 ( \cos(\bk_x) - \cos(\bk_y) )^2$ is
the interplane tunneling matrix element.~\cite{andersen:jpcs:1995:1,ioffe:prb:1998:1}
Note that in this expression $t_\perp(\bk)$ has a strong $\bk$-dependence, with contributions
stemming mainly from the antinodal region of the Brillouin zone.
For convenience, we will express energies in units of the half-bandwidth $D$ of
the electronic dispersion, and the optical conductivity in units
of $\sigma^* = 2 e^2 c t_0^2 / \hbar a b D^2$. In YBa$_2$Cu$_3$O$_y$ compounds,
$D \sim 1eV \sim 8000 \mathrm{cm}^{-1}$ and
$\sigma^*$ is of order $\sigma^*\sim 50 \Omega^{-1} \mathrm{cm}^{-1}$.

In the left panel of Fig.~\ref{fg:sigmac}, we display the computed $\sigma_c(\Omega)$
for three levels of hole-doping and several temperatures. Our results show three distinctive
behaviors.
At high doping $\delta \gtrsim 16\%$, the conductivity displays a metallic-like
behavior, with the build-up of a Drude-like peak as the temperature is decreased.
Note that as the peak increases additional spectral weight appears at
low-energy.
At low doping $\delta \lesssim 10\%$, $\sigma_c(\Omega)$ is characterized by a
gaplike depression at low frequencies where spectral weight is suppressed with
decreasing temperature. The width of the gap when it opens at high temperature
is $\sim 0.15 D$ and remains approximately the same as the temperature is
lowered. Note that the spectral weight that is lost in the gap is
redistributed over a wide range of energies.
The appearance of the depression in the spectra can be directly
linked to the formation of a pseudogap in the antinodal
region.~\cite{ferrero:epl:2009:1,ferrero:prb:2009:1} Indeed, the matrix
element $t_\perp$ appearing in the expression of the optical
conductivity~(\ref{eq:sigmac}) essentially probes the
region~\cite{chakravarty:science:1993:1} close to $(\pm\pi,0), (0,\pm\pi)$ so that a
loss of coherent antinodal quasiparticles results in a loss of low-energy
spectral weight in the $c$-axis optical conductivity. In
Refs.~\onlinecite{ferrero:epl:2009:1,ferrero:prb:2009:1}, it has been shown that in a
zero-temperature analysis of VB-DMFT, coherent quasiparticles disappear in the antinodal
region at a doping $\sim 16\%$. This is consistent with $\sigma_c$ showing a
depression only for doping levels below $\sim 16\%$.
At intermediate doping, $\sigma_c(\Omega)$ has a mixed character (see
left panel of Fig.~\ref{fg:sigmac} for $\delta = 12\%$). Starting from high temperatures, the
low-energy conductivity first displays metallic behavior with an increase in
$\sigma_c$ as the temperature is lowered. At temperatures $T<D/50$, however, a
gap starts to appear and low-energy spectral weight is suppressed.

The right panel of Fig.~\ref{fg:sigmac} displays the experimental interplane
optical conductivity of YBa$_2$Cu$_3$O$_y$ obtained in
Refs.~\onlinecite{schutzmann:prl:1994:1,laforge:prl:2008:1,laforge:prb:2009:1}.
At large doping levels, the data shows a build-up of a Drude peak consistent
with our theoretical calculation. One must note however that the interplane
optical conductivity of YBa$_2$Cu$_3$O$_7$ displays a crossing between the
spectra at different temperatures and that the spectral weight transfer is not
as clear as in the theoretical curves (left panel) for $\delta = 16\%$.  In
fact, only at doping levels larger than $\delta \sim 25\%$ do the computed
$\sigma_c$ eventually behave more metallic-like, with a sharp Drude peak at low
energies.  This behavior is also consistent with other experiments, e.g. on
overdoped YBa$_2$Cu$_3$O$_y$, see Fig.~1 of Ref.~\onlinecite{schutzmann:jpcs:1995:1}.
At low doping, the spectra of YBa$_2$Cu$_3$O$_{6.67}$ display the opening
of a gaplike depression, which is quite well captured by our theoretical results.
Remarkably, $\sigma_c$ is typically a factor of $10$
smaller that in YBa$_2$Cu$_3$O$_{7}$, a factor quantitatively similar to
what we obtain theoretically by comparing our results for $\delta=16\%$
and $\delta=8\%$.
At intermediate doping between these two limits,
YBa$_2$Cu$_3$O$_{6.95}$ displays spectra which depend on temperature
in a non-monotonous manner, first increasing upon cooling and then decreasing at
lower temperatures. This behavior is indeed observed in our theoretical
results at $12\%$ doping. All these
observations show that our calculations agree qualitatively, and to a certain
extent quantitatively, with optical conductivity measurements
on underdoped cuprates (see also e.g.
Ref.~\onlinecite{basov:rmp:2005:1}, Fig.~2 in Ref.~\onlinecite{homes:prl:1993:1}
or Fig.~1 in Ref.~\onlinecite{homes:prb:2005:1}).


\section{Interplane resistivity}

\begin{figure}[ht!]
  \includegraphics[width=8.0cm,clip=true]{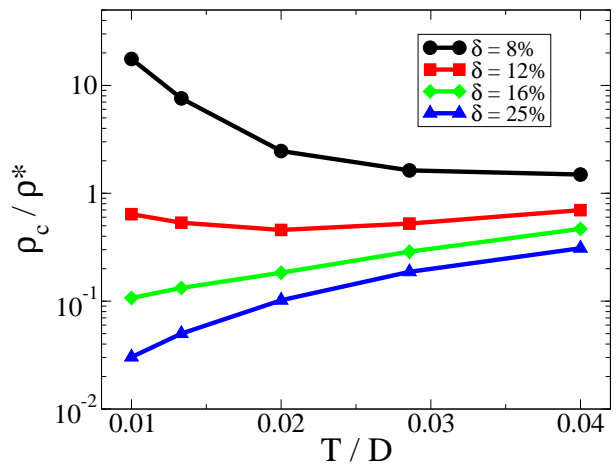}
  \caption{(Color online) The $c$-axis resistivity $\rho_c$
  calculated within the VB-DMFT for four doping levels as a function
  of temperature.}
  \label{fg:rhoc}
\end{figure}

From the $\Omega=0$ extrapolation of the optical conductivity, we obtain the $c$-axis
resistivity $\rho_c = 1/\sigma_c(\Omega=0)$ displayed in Fig.~\ref{fg:rhoc}
for different doping levels, as a function of temperature.
$\rho_c$ is displayed in units of $\rho^* = 1/\sigma^*$
which takes the value $\sim 0.02 \Omega \mathrm{cm}$ in
YBa$_2$Cu$_3$O$_y$. As already discussed for the optical conductivity, the
properties of $\rho_c$ strongly depend on the doping level.
At low doping ($\delta = 8\%$), $\rho_c(T)$ has an insulating behavior with a
large increase as the temperature is lowered. This increase is a direct
consequence of the opening of the pseudogap in the antinodal region of the
Brillouin zone.
%
%
As the doping level is increased (see $\delta=12\%$), $\rho_c(T)$ starts to
display a mixed, non monotonous behavior: it has a minimum
as it changes from metallic-like to insulating-like as $T$ decreases.
This behavior is indeed observed in interplane resistivity measurements on
La$_{2-x}$Sr$_x$CuO$_4$ (Fig.~3 in Ref.~\onlinecite{ito:nature:1991:1}, Fig.~2
in Ref.~\onlinecite{nakamura:prb:1993:1}), YBa$_2$Cu$_3$O$_y$ (Fig.~1 in
Ref.~\onlinecite{takenaka:prb:1994:1}) or Bi$_2$Sr$_2$CaCu$_2$O$_y$ thin films
(inset of Fig.~3 in Ref.~\onlinecite{raffy:pc:2007:1}).
As we go to higher doping, $\rho_c(T)$ eventually acquires a metallic-like
behavior with $d\rho_c/dT > 0$. Note that this metallic behavior emerges from
the transfer of spectral weight to lower energies as can be seen from
Fig.~\ref{fg:sigmac}. Moreover, even at the largest doping level $\delta = 25\%$,
$\rho_c(T)$ does not recover the standard Fermi-liquid behavior with $\rho_c(T)
\sim T^2$ (closer inspection instead shows that $\rho_c(T)$ roughly behaves as
$T^\alpha$ with $\alpha \approx 1.6$). One might need to consider even
larger doping levels to recover a $T^2$ behavior as seen, e.g. in heavily doped
La$_{2-x}$Sr$_x$CuO$_4$.~\cite{ito:nature:1991:1}


\section{Conclusions}

To conclude, in this article we have computed the interplane charge dynamics
within the VB-DMFT framework in order to address the properties of the
normal-state of high-$T_c$ copper-oxide superconductors. Our results show that
the interplane charge dynamics of the normal state is characterized by three regimes.
(i) An overdoped regime where both the $c$-axis optical conductivity and resistivity show a
metallic-like behavior. (ii) An underdoped regime where the destruction of
coherent quasiparticles in the antinodal region induces a gaplike depression in
$\sigma_c$ and a resulting insulating-like interlayer resistivity which
increases as temperature is lowered. (iii) An intermediate-doping mixed regime
with first the buildup of a coherence peak in $\sigma_c$ as the temperature is
decreased, followed by a loss of low-energy spectral weight as the pseudogap
opens at lower temperatures. As a consequence, $\rho_c$ has a non-monotonic
behavior with a minimum at intermediate temperature.  Our calculations
are in good agreement with spectroscopic and transport experiments.


\acknowledgments

We are grateful to H\'el\`ene Raffy for discussions about her experimental
results on $c$-axis resistivity. We acknowledge the support of the Partner
University Fund (PUF-FACE), the Ecole Polytechnique-EDF chair on Sustainable
Energies, the National Science Foundation (NSF-Materials World Network program
and NSF-DMR-0906943), and the Agence Nationale de la Recherche (ANR grant
ECCE). This work was performed using HPC resources from GENCI-CCRT (grant
2009-t2009056112). We also acknowledge the Kavli Institute for Theoretical
Physics where part of this work was completed (under grant NSF-PHY05-51164).
DNB is supported by the NSF.

\textit{Note added:} After initial submission of this manuscript, Lin, Gull and
Millis (arXiv:1004.2999) reported calculations of c-axis optical conductivity
in 8-site cluster DMFT. When comparison is possible, consistency between these
results and our VB-DMFT (2-site) results is observed.


\appendix

\section{Details about the VB-DMFT approach}
\label{app:vbdmft}

Let us give here some technical details and briefly describe how our results were
obtained. Within the VB-DMFT, the effective model that one has to solve is a
two-impurity Anderson model~\cite{ferrero:epl:2009:1,ferrero:prb:2009:1}. The
even (+) combination of the two impurity orbitals is associated with the nodal
part of the Brillouin zone, while the odd (-) combination is associated to the
antinode. The Anderson impurity model is solved using a continuous-time quantum
Monte Carlo algorithm (CTQMC)~\cite{werner:prl:2006:1,werner:prb:2006:1}
that yields the even (resp. odd) orbital self-energy $\Sigma_+(i\omega_n)$ (resp.
$\Sigma_-(i\omega_n)$) at the Matsubara frequencies $i\omega_n$.
At this stage one can readily get the quasiparticle residues
by carefully extrapolating $\Sigma_\pm(i\omega_n)$ at zero frequency
\begin{equation}
  Z_\pm = \Big( 1 - \frac{d\mathrm{Im}\Sigma_\pm(i\omega_n)}{d\omega_n}
          \Big|_{\omega_n \rightarrow 0} \Big)^{-1}.
\end{equation}
The inverse quasiparticle lifetime is then given by
\begin{equation}
  \Gamma_\pm = - Z_\pm \mathrm{Im} \Sigma_\pm(i\omega_n) \Big|_{\omega_n \rightarrow 0}.
\end{equation}
Finally, the nodal ($\Gamma_\mathrm{N}$) and antinodal ($\Gamma_\mathrm{AN}$) inverse lifetimes
are associated with $\Gamma_+$ and $\Gamma_-$ respectively.
In order to compute the optical conductivity and the resistivities it is
necessary to have access to the in-plane spectral function $A(\bk,\omega)$.
A first step is to analytically continue the self-energies $\Sigma_\pm$ on the
real-frequency axis, which is achieved using Pad\'e
approximants.~\cite{vidberg:jltp:1977:1} This is made possible thank to the
extremely accurate, low-noise imaginary-time data.
A second step is to interpolate the self-energy to obtain a $\bk$-dependent
$\Sigma(\bk,\omega)$ from which the spectral function will be computed.
Refs.~\onlinecite{ferrero:epl:2009:1,ferrero:prb:2009:1} have shown that an
efficient interpolation procedure is
the $M$-interpolation.~\cite{stanescu:prb:2006:1,stanescu:ap:2006:1}
The self-energy is obtained with
$\Sigma(\bk,\omega) = \omega + \mu - M(\bk,\omega)^{-1}$, where
the cumulant $M(\bk,\omega)$ is given by
\begin{equation}
   M(\bk,\omega) = \frac{\alpha_+(\bk)}{\omega+\mu-\Sigma_{+}(\omega)} +
   \frac{\alpha_-(\bk)}{\omega+\mu-\Sigma_{-}(\omega)},
\end{equation}
with $\alpha_\pm(\bk) = \frac{1}{2}\{1\pm\frac{1}{2}[\cos(\bk_x)+\cos(\bk_y)]\}$.
Finally, the spectral function is computed with
\begin{equation}
  A(\bk,\omega) = \frac{1}{\omega + \mu - \epsilon_\bk - \Sigma(\bk,\omega)},
\end{equation}
where $\epsilon_\bk$ is the square lattice dispersion.


\bibliography{caxisdynamics}

\end{document}